\documentclass[a4paper,10pt]{article}
\usepackage{amsfonts,amssymb,amsmath}
\usepackage{algorithm,algorithmic}
\usepackage{natbib}
    \newcommand{\mvbar}{\middle\vert}
	\newcommand{\ud}{\,\mathrm{d}}
	\newcommand{\U}{\text{U}}
	
\title{Contribution to the discussion of ``Sequential Quasi-Monte-Carlo Sampling'' -- Gerber, Chopin}
\author{M. Pollock, A. M. Johansen, K. {\L}atuszy\'{n}ski, G. O. Roberts\\ Dept. of Statistics, University of Warwick, UK, CV4 7AL \date{}}

\begin{document}
\maketitle
\setcounter{tocdepth}{5}

\noindent We congratulate the authors on an excellent paper. It has inspired
us to consider ways to incorporate QMC within
SMC schemes in settings in which the transition density of the latent process
is intractable and pseudo-marginal methods are deployed. In particular,
consider filtering for partially observed (jump)
diffusions (e.g. \cite{JRSSB:FPR08,PhD:P13,B:PJR14}), in which (in the simplest
setting) the latent process is a diffusion satisfying the SDE, 
\begin{align}
\ud X_t = \alpha(X_t)\ud t + \ud B_t,\quad X_0=x, \quad t\in[0,T].
\end{align}
In this setting the data comprise partial observations (arising at a finite
collection of time points) of the latent process; the extension to noisy
observation is trivial. The transition density of the latent process (under
certain regularity conditions) can be shown \citep{S:DF86} to have the
following form between any two time points $a$ and $b$, where $0\leq a<b\leq T$,
with $\mathbb{W}^{x_a,x_b}_{a,b}$ denoting the law of a Brownian bridge between
$x_a$ and $x_b$ over $[a,b]$: 
\begin{align}
  p_{b-a}(x_b|x_a) = &\notag
  \underbrace{\mathcal{N}(x_b;x_a,b-a)\exp\left\{\int^{x_b}_{x_a} \alpha(u)\ud
      u\right\}}_{\propto \tilde{p}_{b-a}(x_b|x_a)}\\
& \cdot \mathbb{E}_{\mathbb{W}^{x_a,x_b}_{a,b}} \underbrace{\left[\exp\left\{-\int^b_a \dfrac{\alpha^2(W_t)+ \alpha^\prime(W_t)}{2}\ud t \right\}\right]}_{=:\psi(W)}.
\end{align}
To propagate particles between consecutive observation times, $a$ and $b$, one
could simulate from the proposal $\tilde{p}_{b-a}(x_b|x_a)$, perhaps by
rejection sampling, and modify the weight of each particle by a factor
corresponding to an unbiased estimate of $\psi(W)$ (where $W\sim\mathbb{W}^{x_a,x_b}_{a,b}$). Supposing that 
\[\forall t\in[0,T]\qquad \left(\alpha^2(W_t)+\alpha^\prime(W_t)\right)/2\in[L,U],\] 
and letting $\kappa \sim \textrm{Poi}\left[(U-L)\cdot(b-a)\right]$ and
$(\xi_1,\ldots{})\overset{\text{iid}}{\sim} \U[a,b]$, we have the representation:
\begin{align}
\psi(W) & = e^{-L\cdot(b-a)}\cdot\mathbb{E}\left[\mathbb{E}\left[\prod^\kappa_{i=1}\dfrac{2U- \alpha^2(W_{\xi_i})-\alpha^\prime(W_{\xi_i})}{2(U-L)}\,\mvbar\, \kappa, W \right]\,\mvbar\, W\right].
\end{align}
An unbiased estimate of $\psi(W)$, using a finite dimensional
realisation of the sample path, is obtained by sampling $\kappa$ and $\xi_{1},\ldots,\xi_{\kappa}
\overset{\text{iid}}{\sim} \U[a,b]$ and employing a simple Monte Carlo
approximation \citep{JRSSB:BPRF06}.\\
\\
This scheme uses an unbiased estimator constructed by simulating $\kappa$ and then using a $\kappa$-dimensional uniform random variable to approximate the inner
expectation. Finding a lower variance unbiased estimator of $\psi(W)$ is
desirable, and one would like to exploit RQMC. However, the
dimension of the random variable being random, it is not straightforward to do
this directly. One \emph{could} instead sample $\kappa$ in the usual manner, and
approximate the inner expectation conditionally using an RQMC point set.
There is clearly a computational cost associated with such a RQMC method, which will only be appropriate for problems in which the variance of the
simple Monte Carlo estimator of $\psi(W)$ is large and $\kappa$ is typically small.
\bibliography{BibtexReferences}

\end{document}